\documentclass[twocolumn]{aastex6}
\begin{document}
\title{Phosphorus Abundances in FGK Stars}
\author{Z. G. Maas}
\affil{Indiana University Bloomington, Astronomy Department, Swain West 319, 727 East Third Street, Bloomington, IN 47405-7105, 
USA}
\email{zmaas@indiana.edu}

\and

\author{C. A. Pilachowski} 
\affil{Astronomy Department, Indiana University Bloomington, Swain West 319, 727 East Third Street, Bloomington, IN 47405-7105, 
USA}
\email{cpilacho@indiana.edu}

\author{G. Cescutti}
\affil{I.N.A.F. Osservatorio Astronomico di Trieste, via G.B. Tiepolo 11, 34131, Trieste, Italy}
\email{cescutti@oats.inaf.it}
\begin{abstract}

We measured phosphorus abundances in 22 FGK dwarfs and giants that span --0.55 $<$ [Fe/H] $<$ 0.2 using spectra obtained with the Phoenix high resolution infrared spectrometer on the Kitt Peak National Observatory Mayall 4m telescope, the Gemini South Telescope, and the Arcturus spectral atlas. We fit synthetic spectra to the \ion{P}{1} feature at 10581 $\AA$ to determine abundances for our sample. Our results are consistent with previously measured phosphorus abundances; the average [P/Fe] ratio measured in [Fe/H] bins of 0.2 dex for our stars are within $\sim$ 1 $\sigma$  compared to averages from other IR phosphorus studies. Our study provides more evidence that models of chemical evolution using the results of theoretical yields are under producing phosphorus compared to the observed abundances. Our data better fit a chemical evolution model with phosphorus yields increased by a factor of 2.75 compared to models with unadjusted yields. We also found average [P/Si] = 0.02 $\pm$ 0.07 and [P/S] = 0.15 $\pm$ 0.15 for our sample, showing no significant deviations from the solar ratios for [P/Si] and [P/S] ratios.

\end{abstract}


\section{Introduction}

Abundance measurements of the light elements, from carbon to argon, are important in the study of nucleosynthesis and galactic chemical evolution. The light elements have been used to study differences in Galactic structure, multiple populations in stellar clusters, and used as proxies for metallicity in extragalactic studies. The even elements in particular, including O, Mg, Si, S, and Ca, and their nucleosynthesis in massive stars has been studied in detail \citep{nomoto13}. These even elements are well understood in the context of hydrostatic burning and their yields, coupled with chemical evolution models, successfully predict observed abundance trends (e.g. \citealt{nomoto13}). However, the odd light elements are thought to be produced via other processes and few abundance measurements exist for elements such as P and Cl in stars \citep{nomoto13}. This work focuses on the odd element phosphorus, an important element for life \citep{macia} with still uncertain nucleosynthesis production mechanisms.  

Phosphorus has only one stable isotope, $^{31}$P, and is thought to be produced mostly in massive stars through neutron capture on Si isotopes, specifically $^{30}$Si, in hydrostatic carbon and neon burning shells \citep{woosley95}. Phosphorus abundances have been observed using atomic infrared lines at approximately 10500-10820 $\AA$ and atomic lines in the near-UV at 2135-2136 $\AA$. Phosphorus abundances have been derived in planetary nebulae using \ion{P}{3} lines located at $\sim$ 7875 $\AA$ \citep{pottasch08,otsuka11} and in damped Lyman alpha systems using ionized phosphorus lines, such as \ion{P}{2} line at 1152 $\AA$ \citep{outram99,molaro01,levshakov02}. Anomalously high phosphorus abundances have been measured using optical phosphorus features in blue horizontal branch stars \citep{behr99}, which may be due to diffusion of heavy elements, including P, in the photosphere \citep{michaud08}. Molecular forms of phosphorus, such as PO, PN, and CP, have been detected and used to understand phosphorus chemistry in the interstellar medium using features at millimeter and radio wavelengths. For example, phosphorus molecules have been detected in the interstellar medium \citep{turner87} and in star forming regions \citep{fontani16,lefloch16}. Phosphorus molecules have also been found in the circumstellar envelopes of evolved stars \citep{milam,agundez14}. Finally, the diffuse interstellar medium has been measured using \ion{P}{2} lines at 1125 $\AA$ and 1533 $\AA$ \citep{lebouteiller05}.
 
Solar phosphorus measurements of the infrared lines using 3-D solar models found an abundance of A(P)=5.46 $\pm$ 0.04 \citep{caffau07}. The infrared lines have been used to determine phosphorus abundances in F and G spectral type stars between --1.0 $<$ [Fe/H] $<$ 0.2 \citep{caffau11,caffau16}. The P abundance differences in solar twins were examined by \citet{melendez09} and the P abundance of Procyon was determined by \citet{kato96}. These lines weaken for stars with low abundance and are typically only observed in stars with [Fe/H] $\gtrsim$ --1.0. The near-UV phosphorus lines however can be detected and measured in metal poor stars \citep{roederer14}. These features have been used to measure P abundances in FKG type stars with metallicities between --3.3 $<$ [Fe/H] $<$ --0.2 \citep{jacobson14,roederer14,spite17}. Phosphorus abundances have also been compared to alpha elements; \citet{caffau11} found a constant phosphorus to sulfur ratio of [P/S]= 0.10 $\pm$ 0.10 over their metallicity range. The odd, light elements, such as Na, Al, and P are thought to be produced through neutron capture and their abundances should be sensitive to the neutron flux in massive stars during hydrostatic shell burning \citep{woosley95}. The neutron flux is dependent on metallicity and a decreasing ratio of phosphorus to elements not dependent on neutron flux, such as the alpha elements is expected. The constant [P/S] ratio found by \citet{caffau11} suggests P production is insensitive to the neutron excess and other processes may be important in the nucleosynthesis of phosphorus. This effect is more pronounced at lower metallicities of [Fe/H] $<$ --1.0 and more measurements would confirm the constancy of the [P/S] ratio with metallicity \citep{caffau11}. 

Current models of phosphorus production do not match the observed abundances. Chemical evolution models currently predict subsolar phosphorus abundances at [Fe/H] = 0 for the solar neighborhood \citep{kobayashi11}. A possible exception is the result obtained by \citet{gibson}. Phosphorus yields would have to be increased by factors of 1.5-3 to match the observed [P/Fe] ratios measured in field stars \citep{cescutti12}. Additional P production mechanisms have been proposed, such as $\alpha$-particle capture on $^{27}$Al or proton capture on $^{30}$Si, to resolve the abundance discrepancy \citep{caffau11}.

Additional abundance measurements are therefore necessary to help resolve the issue of the nucleosynthesis of phosphorus. It is uncertain how much yields must be increased to match models and additional measurements are needed to understand if the [P/Fe] versus metallicty slope of the chemical evolution model accurately fits the data. We measured phosphorus in 22 FGK field dwarfs and giants with metallicities in the range --0.5 $\lesssim$ [Fe/H] $\lesssim$ 0.2 to test these questions. Section \ref{section::datared} describes the data reduction. The methodology used to determine phosphorus abundances is discussed in section \ref{section::abundance}. The results are compared to chemical evolution models in section \ref{section::discussion} and the final conclusions are summarized in section \ref{section::conclusion}. 

\section{Observations and Data Reduction} \label{section::datared} 

Our sample consists of 19 stars observed using the high resolution infrared spectrometer Phoenix on the Kitt Peak National Observatory 4m Mayall telescope and two stars from proposal proposal ID GS-2016B-Q-76 on Gemini South Telescope, also observed using Phoenix. We also measured the phosphorus abundance of Arcturus using the infrared atlas \citep{hinkle95}. The full list of stars observed is found in Table \ref{table::obslog}. Target stars with known atmospheric parameters were chosen because the available wavelength range is narrow (our spectral range spanned only $\sim$ 50 $\AA$) and contained too few spectral features to determine atmospheric parameters independently. Also, the targeted \ion{P}{1} features are strongest in stars with effective temperatures between $\sim$4500 K and $\sim$7000 K. Our dwarf star sample was selected from \citet{bensby14}, \citet{reddy03}, and \citet{ramirez13}. Atmospheric parameters for the sample of dwarf stars were adopted from \citet{ramirez13,bensby14} and additional abundances for Si and S are available from \citet{reddy03} for a portion of our sample of stars. Additional giant stars with appropriate atmospheric parameters were chosen from the Pascal catalogue \citep{soubiran}. For all stars, the atmospheric parameters were determined spectroscopically for each literature source (given in Table \ref{table::params}). Finally, only targets with bright J-magnitudes (J $\lesssim$ 7 mags), obtained from 2MASS \citep{skrutskie06}, were chosen from for observations to ensure sufficient signal to noise ratio to measure weak phosphorus features.

\begin{deluxetable}{ c c c c c c}
\tablewidth{0pt} 
\tabletypesize{\footnotesize}
\tablecaption{Summary of Phoenix Observations \label{table::obslog}} 
 \tablehead{\colhead{HD} & \colhead{UT Date} & \colhead{Telescope}& \colhead{Spectral\tablenotemark{a}} & \colhead{J\tablenotemark{b}} & \colhead{S/N} \\
 \colhead{Number} & \colhead{} & \colhead{} & \colhead{Type} & \colhead{(Mag)} }
\startdata
20794  & 2016 Dec 15 & 1  &  G8III   &  3.032  & 280 \\
46114  & 2016 Dec 15 & 1 & G8V    & 6.266 & 250 \\
107950 & 2015 June 3  & 2    & G6III & 3.476 & 110 \\
 120136 & 2015 June 6 & 2&    F6IV          &   3.620 & 170 \\
 121560 & 2015 June 3 & 2&   F6V & 5.137      & 160 \\
 124819 & 2015 June 6  & 2& F5    &   6.610 & 160 \\
 126053 & 2015 June 3  & 2& G1.5V   &   5.053 & 140 \\
 136925 & 2015 June 4 & 2& G0 &  6.751 & 140 \\
 140324 & 2015 June 3 & 2& G0IV/V &  6.295 & 80 \\
 148049 & 2015 June 4 & 2& F8 &  6.355 & 170 \\
 151101 & 2015 June 6 & 2& K0III &  2.913 & 170 \\
 152449 & 2015 June 3 & 2& F6V  &  6.823 & 60 \\
 160507 & 2015 June 6 & 2& G5III &  5.169 & 170 \\
 163363 & 2015 June 3 & 2&  F8 &  6.788 & 70 \\
 167588 & 2015 June 4 & 2&  F8V &  5.363 & 210 \\
 174160 & 2015 June 4 & 2& F7V &  5.244 & 270 \\
 186379 & 2015 June 4 & 2& F8V & 5.741 & 190 \\
 186408 & 2015 June 3 & 2& G1.5Vb &   5.090 & 190 \\
 191649 & 2015 June 3 & 2& G0 &  6.323 & 150 \\
 193664 & 2015 June 6 & 2& G3V &  4.879 & 210 \\
 194497 & 2015 June 4 & 2& F6V &  6.504 & 120\\
\enddata
\tablenotetext{a}{spectral types from the SIMBAD database}
\tablenotetext{b}{J magnitudes from 2MASS \citep{skrutskie06}}
\tablecomments{Telescopes (1) Gemini South Telescope;\\ (2) KPNO Mayall 4m Telescope}
\end{deluxetable}

The program stars were observed with the Phoenix infrared spectrometer \citep{hinkle_et_al_1998} at the f/16 focus of the KPNO Mayall 4 meter telescope in 2015 June and with Phoenix on Gemini South in December 2016. The 0.7 arcsecond, four-pixel slit was used, resulting in a spectral resolution of $\sim$50,000 with both telescopes. Echelle order 53 was selected with a narrow band order sorting filter to observe the wavelength range 1.0570 $\mu$m - 1.0620 $\mu$m. Standard observing procedures for infrared observations were followed \citep{joyce1992}; each object was nodded between two slit positions in an 'ABBA' pattern. Dark and flat field images were observed at the beginning of each night.

The data reduction was accomplished using the IRAF software suite\footnote{IRAF is distributed by the National Optical Astronomy Observatory, which is operated by the Association of Universities for Research in Astronomy, Inc., under cooperative agreement with the National Science Foundation.”}.  The dark images were combined using a median filter with a sigclip rejection algorithm and the flat images were median combined with an avsigclip rejection algorithm. The combined dark image was subtracted from the combined flat image. Sky contribution, bias, and detector blemishes were removed by subtracting objects from one nod position against the next position ('A' slit position image subtracted from the 'B' slit position image and visa-versa). After sky subtraction, each object was flatfielded using a dark corrected, normalized flatfield image. Spectral type A and B standard stars were observed for telluric line removal, however we found no telluric lines in our spectral region. An atlas of Arcturus also shows no telluric contamination in this spectral range, therefore no telluric correction was needed \citep{hinkle95}. Wavelength calibrations were done using the stellar spectral lines, as few comparison lamp emission lines are available in our narrow wavelength range.

\section{Abundance Analysis}
 \label{section::abundance}
 
 \subsection{Abundance Analysis}
Abundances were obtained using MOOG spectral synthesis software \citep{sneden} (Version 2014) with MARCS model atmospheres \citep{gustafsson}. Atmospheric parameters were taken from the literature; a full list of atmospheric parameters and their sources is found in Table \ref{table::params}. Additionally, the synthetic spectrum of HD 120136 was broadened to reflect the rotational velocity of vsini = 15.4 km/s \citep{mallik03}.

\begin{deluxetable*}{l c c c c c l l l l}
\tabletypesize{\footnotesize}
\centering
\tablewidth{0pt} 
\tablecaption{Atmospheric Parameters and Abundances \label{table::params}} 
\tablehead{ \colhead{Star} & \colhead{T$_{eff}$ } & \colhead{log g} & \colhead{[Fe/H]} & \colhead{$\xi$ } & \colhead{Ref.} & \colhead{A(Si)} & \colhead{A(P)} & \colhead{A(S)\tablenotemark{a}} & \colhead{[P/Fe]} \\ 
\colhead{HD Number} & \colhead{(K)} & \colhead{} & \colhead{} & \colhead{(km s$^{-1}$)} } 
\startdata
20794  &	5372&   	4.50	&	--0.46&	0.79	&	2	&	7.30 $\pm$ 0.08 	&	5.36 $\pm$ 0.17	& \nodata &	0.36	\\
46114   &	 5129 & 3.50      &	--0.44 &	0.95	&	2	&	7.17 $\pm$ 0.06 	&	5.04 $\pm$ 0.09	& \nodata & 0.02	\\
107950	&	5100&   	2.50	&	--0.13&	1.7	&	3	&	7.50 $\pm$ 0.07 	&	5.39 $\pm$ 0.08	&	\nodata & 0.11	\\
120136	&	6339	&	4.19	&	0.23	&	1.7	&	4	&	7.75	 $\pm$ 0.07 &	5.68 $\pm$ 0.09 &	\nodata &--0.01	\\
121560	&	6139	&	4.29	&	--0.39	&	1.36	&	5	&	7.09 $\pm$ 0.06	&	5.14 $\pm$ 0.04	&	7.00 &0.07	\\
124819	&	6174	&	4.21	&	--0.27	&	1.55	&	5	&	7.32 $\pm$ 0.06	&	5.29	 $\pm $ 0.06&	7.17 &0.10\\
124897\tablenotemark{b} & 4286 & 1.66 & --0.52 & 1.75 & 1 & 7.32 $\pm$ 0.04\tablenotemark{c} & 5.11 $\pm$ 0.15 & \nodata &0.22 \\
126053  &   5691 &   4.44    &   --0.36  &   1.07 & 5   &  7.15 $\pm$ 0.06 & 5.11 $\pm$ 0.06 &  \nodata &0.06 \\ 
136925	&	5732	&	4.32	&	--0.29	&	0.98	&	5	&	7.33 $\pm$ 0.08	&	5.37 $\pm$ 0.07 	&	7.24 &0.11	\\
140324	&	5913	&	4.05	&	--0.3	&	1.27	&	5	&	7.24 $\pm$ 0.07	&	5.31	 $\pm$ 0.08 &	7.14 &0.03	\\
148049	&	6071	&	4.16	&	--0.33	&	1.37	&	5	&	7.28	 $\pm$ 0.06 &	5.26	 $\pm$ 0.06 &	7.14 &0.13	\\
151101	&	4535	&	2.1	&	--0.14	&	2.28	&	3	&	7.45 $\pm$ 0.11	&	5.36	 $\pm$ 0.12 &	\nodata &0.04	 \\
152449	&	6154	&	4.15	&	--0.03	&	1.48	&	5	&	7.55 $\pm$ 0.07	&	5.46 $\pm$ 0.08	&	7.41 &0.03\\
160507	&	4790	&	2.648	&	--0.23	&	1.57	&	6	&	7.47	 $\pm$ 0.10 &	5.49	 $\pm$ 0.13 &\nodata &	0.26\\
163363	&	5970	&	3.89	&	--0.04	&	1.48	&	5	&	7.47 $\pm$ 0.07 	&	5.55	 $\pm$ 0.08 & 7.38 &0.13		\\
167588	&	5845	&	3.92	&	--0.39	&	1.39	&	5	&	7.22	 $\pm$ 0.05 &	5.22	 $\pm$ 0.06 &	7.08 &0.15	\\
174160	&	6417	&	4.36	&	--0.01	&	1.5	&	5	&	7.4 $\pm$ 0.05	&	5.39	 $\pm$ 0.04 &	7.28 &--0.06	\\
186379	&	5899	&	4.02	&	--0.37	&	1.4	&	5	&	7.21 $\pm$ 0.05	&	5.19	 $\pm$ 0.04 &	7.08 &0.10	\\
186408	&	5787	&	4.26	&	0.05	&	1.21	&	5	&	7.65 $\pm$ 0.06	&	5.54 $\pm$ 0.05	&	7.55 &0.03\\
191649	&	6081	&	4.26	&	--0.2	&	1.46	&	5	&	7.29 $\pm$ 0.06	&	5.26 $\pm$ 0.06	&	7.28 &0	\\
193664	&	5915	&	4.43	&	--0.13	&	1.16	&	5	&	7.4 $\pm$ 0.06	&	5.24 $\pm$ 0.05	&	7.33 &--0.09	 \\
194497	&	6198	&	3.78	&	--0.43	&	1.89	&	5	&	7.18 $\pm$  0.06	 &	5.14 $\pm$ 0.05 & 	7.05 &0.11	\\
\enddata
\tablenotetext{a}{S abundances from \citet{reddy03}}
\tablenotetext{b}{$\alpha$ Boo}
\tablenotetext{c}{Si abundance from \citet{ramirez11}}
\tablecomments{References: (1) \citealt{ramirez11}; (2) \citealt{bensby14}; (3) \citealt{hekker}; (4) \citealt{santos}; (5) \citealt{ramirez13}; (6) \citealt{wang} }
\end{deluxetable*}

\begin{figure*}[t!]
\epsscale{1} 
\centering
\includegraphics[trim=5cm 5cm 0cm 0cm, scale=.25, clip=True]{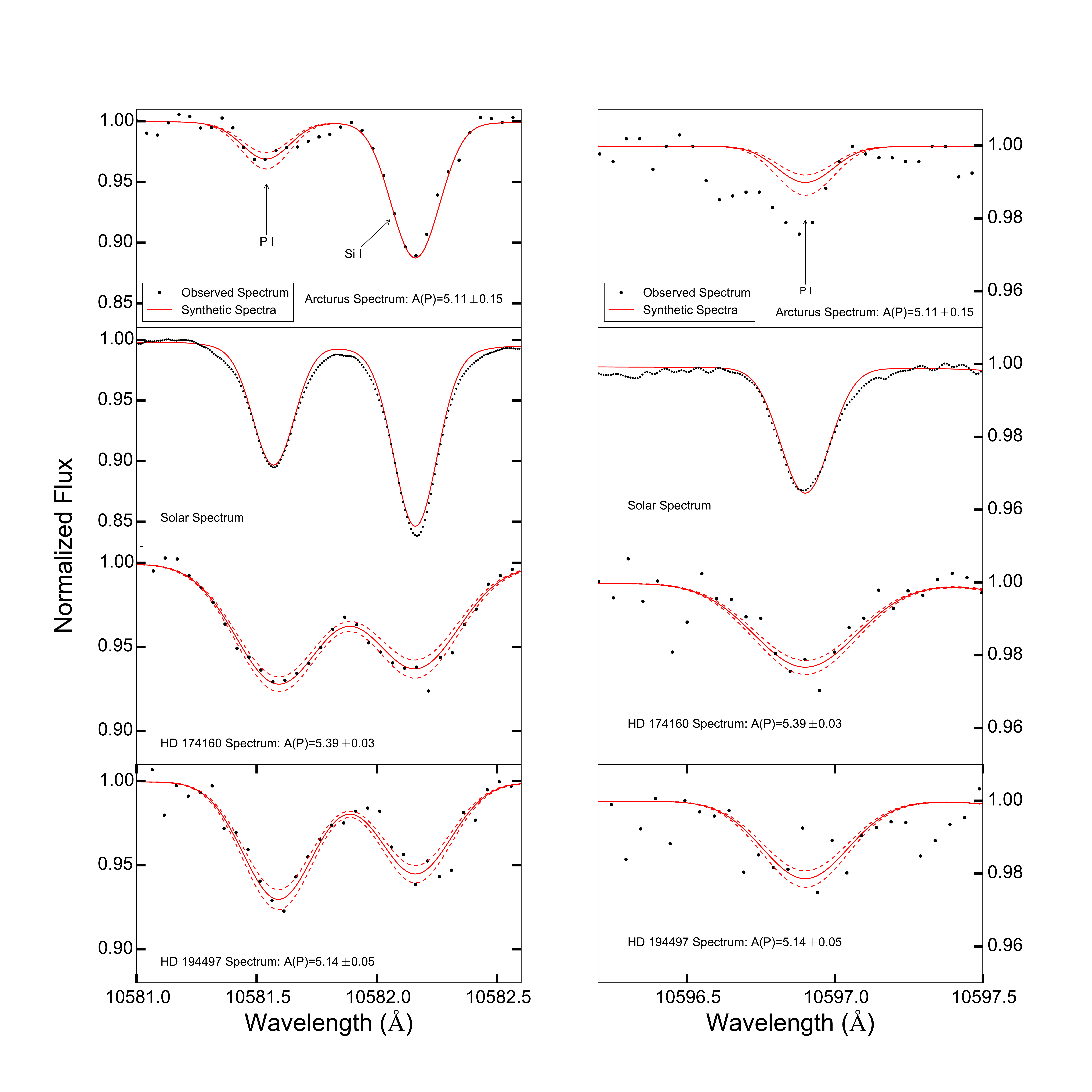} 
\caption{Each panel shows observed spectra (black circles) and the synthetic fits to the spectra (red lines). The solid red lines represent the best fit while the dashed lines repesent the uncertainty on only the fit.  \label{fig::spectra}} 
\end{figure*}

Atomic line data for the phosphorus transitions shown in Table \ref{table::lines}, was obtained from \citet{berzinsh} and is identical to that used in \citet{caffau11}. Atomic line data for a feature \ion{Si}{1} was taken from the Kurucz database\footnote{http://kurucz.harvard.edu}. The gf value for the \ion{Si}{1} line at 10582 $\AA$ was refined by fitting the solar spectrum obtained from the \citet{wallace93} atlas and the infrared Arcturus atlas \citep{hinkle95}. We adopted solar atmospheric  parameters of T= 5870 K, log g= 4.44, and a microturbulence of 0.75 kms$^{-1}$. Atmospheric parameters for Arcturus were adopted from \citet{ramirez11}. Silicon abundances for the solar spectrum of A(Si) = 7.51 and A(Fe) = 7.50 are adopted from \citet{asplund09}, the silicon abundance for Arcturus is from \citet{ramirez11}, the solar phosphorus abundance of A(P) = 5.46 is adopted from \citet{caffau07}, and the solar sulfur abundance of A(S) = 7.16 is from \citet{caffau11sulfur}. The solar spectrum synthesis was performed using the atomic data from Table \ref{table::lines} and overplotted on the solar spectrum in Figure \ref{fig::spectra}. The best fit was determined by eye and the log gf value for the Si line was increased by 0.07 to fit both the solar and Arcturus spectra. The updated log gf value for the \ion{Si}{1} line listed in Table \ref{table::lines}.

\begin{deluxetable}{l l l l l}
\tabletypesize{\small}
\centering
\tablewidth{0pt} 
\tablecaption{Line List \label{table::lines}} 
\tablehead{ \colhead{Element} & \colhead{$\lambda$ $_{air}$} &  \colhead{$\chi$} & \colhead{log gf} & \colhead{Transition} \\
\colhead{} & \colhead{$\AA$} & \colhead{ (eV)} & \colhead{} & \colhead{} } 
\startdata
P I & 10581.569 & 6.99 & 0.45 & 4s$^{4}$P$_{5/2}$--4p$^{4}$D$^{0}_{7/2}$ \\
Si I & 10582.16 & 6.222 & -1.099 & 4p$^{1}$D$_{2}$--6s$^{1}$P$_{1}$ \\
P I & 10596.900 & 6.940 & --0.21 &  4s$^{4}$P$_{1/2}$--4p$^{4}$D$^{0}_{1/2}$ \\
\enddata
\end{deluxetable}

Phosphorus abundances were determined by comparing synthetic spectra to the observed spectra and determining the best fit by eye. Examples of final fits to the spectra are shown in Figure \ref{fig::spectra}. To test the reliability of abundances determined by eye, abundances were re-determined by minimizing the $\chi^{2}$ between synthetic spectra and the corresponding observation. We found the abundances determined from the $\chi^{2}$ minimization technique were consistent with those found by eye; the average difference was 0.01 $\pm$ 0.05 dex. The largest outlier in this difference is the star HD 163363 which has a difference of --0.12 dex from their $\chi^{2}$ minimization abundance subtracted from their by eye abundance. This star has a low signal to noise of 70 and this may be the cause of the discrepancy.

The phosphorus abundance determined from the 10581 $\AA$ line was chosen as the final phosphorus abundance because the other feature at 10596 $\AA$ was often too weak to determine reliable abundances. For dwarf stars with high S/N measurements, such as HD 174160 (in Fig \ref{fig::spectra}), HD 193664, HD 186379, and HD 167588, the P abundance from the 10596 $\AA$ agreed with the \ion{P}{1} feature at 10581 $\AA$. However, the fit to the \ion{P}{1} feature at 10596 $\AA$ does not match the the observed Arcturus spectrum in Figure \ref{fig::spectra}. This is likely due to blend with a weak, unidentified line in cool giants. Final phosphorus abundances are plotted in Figure \ref{fig::phos_evolution}. We find no significant dependence of the phosphorus abundance on temperature for our stars, also shown in Figure \ref{fig::phos_evolution}. 

\subsection{Abundance Uncertainty}

Uncertainties in the phosphorus abundances were determined in two ways; first from uncertainties in the atmospheric parameters and from the fit to the phosphorus features. The uncertainty on the abundance was computed by determining the range of acceptable models by eye; the uncertainty determined this way was typically 0.05. This is consistent with the uncertainty in the $\chi^{2}$ minimization determined from the abundance difference at the p = 0.05 significance level, typically 0.07 dex.

To calculate the uncertainties due to the atmospheric parameters, we re-computed the abundances, varying the model atmosphere considering these uncertainties, fit synthetic spectra to the observed spectra to re-derive abundances, and added all errors found from the differences between the abundance in Table \ref{table::params} and those computed with the varied atmospheric models, from each atmospheric parameter term in quadrature. The average uncertainty in our model parameters in the dwarf stars stars from \citet{ramirez13} were $\delta$T= 49 $\pm$ 14, $\delta$log g= 0.04 $\pm$ 0.01, $\delta$[Fe/H]= 0.05 $\pm$ 0.01, and $\delta \xi$= 0.12 kms$^{-1}$. Due to the uniformity of the uncertainty conditions, uncertainties were calculated for three stars from \citet{ramirez13} and final uncertainties found from fitting synthetic spectra were 0.03 dex for temperature, and an uncertainty in abundance of 0.01 due to [Fe/H] and 0.01 dex from the microturbulence. The abundance uncertainty from varying the log g was consistent with zero dex. Added in quadrature, the total uncertainty due to the atmospheric parameters in phosphorus abundance for stars from \citet{ramirez13} was 0.03 dex. 

Stars with atmospheric parameters not from \citet{ramirez13} had larger uncertainties in their atmospheric parameter with averages of $\delta$T= 85 $\pm$ 11, $\delta$log g= 0.17 $\pm$ 0.06, $\delta$[Fe/H]= 0.09 $\pm$ 0.01, and $\delta \xi$= 0.20 kms$^{-1}$ \citep{hekker,wang,santos}. The uncertainties from the atmospheric parameters were calculated for each of these stars individually and average uncertainties were found to be 0.08 dex for temperature, 0.04 dex for log g, 0.02 for [Fe/H] changes, and the microturbulence caused an uncertainty 0.01. For all stars, the uncertainty were added together in quadrature with uncertainties in the fit and the final total uncertainties are found in Table \ref{table::params}. The uncertainties on [P/Fe] include both the uncertainty on the phosphorus abundance and on [Fe/H]; $\pm$ 0.05 dex uncertainty for stars from \citet{ramirez13} and $\pm$ 0.09 dex uncertainty from other sources \citep{hekker,wang,santos}, as listed in Table \ref{table::params}.

Our abundance determinations were also performed with the assumptions of LTE and 1-D MARCS atmospheric models, plane parallel models for dwarf stars and spherical models for giants. No study of NLTE effects on the phosphorus lines is available, however \citet{asplund09} suggests that NLTE effects should be minimal because the phosphorus lines are expected to behave similarly to \ion{S}{1} lines with similar transitions. Specifically, LTE is approximately valid for the 8693 $\AA$ (4p$^{5}$P$_{3}$--4d$^{5}$D$^{0}_{3}$) and 8694 $\AA$ lines \ion{S}{1} lines (4p$^{5}$P$_{3}$--4d$^{5}$D$^{0}_{4}$) \citep{chen02, takeda05}. The effects of 3-D stellar models compared to 1-D were $\sim$ 0.03 when calculating the phosphorus abundance for the sun \citep{caffau07}. 

\subsection{Literature Comparisons}

Our silicon abundances can be compared to Si measurements previously derived in 15 stars in common with \citet{reddy03}. \citet{reddy03} had derived different metallicities than \citet{ramirez13} and therefore reported [Si/Fe] values reflected both differences in Si and Fe abundances between each sample. The differences in metallicity were removed between the two studies and the A(Si) values were compared directly. The difference between our measurements and those from \citet{reddy03} is --0.03 $\pm$ 0.11 dex indicating agreement with scatter. Finally, the star HD 120136 was also studied by \citet{caffau11} and our result of [P/Fe] = --0.01 $\pm$ 0.09 is consistent with their measurement, [P/Fe]= 0.03 $\pm$ 0.08. 

The two stars, HD 20794 and HD 46114 are consistent with silicon abundances derived by \citet{bensby14}. Our abundance for HD 20794 is [Si/Fe] = 0.25 $\pm$ 0.12 and the abundance for HD 46114 is 0.1 $\pm$ 0.08. The abundance from \citet{bensby14} for HD 20794 is   0.22 $\pm$ 0.22 and the abundance for HD 46114 is 0.09 $\pm$ 0.06.  
\begin{deluxetable}{c c c c}
\tabletypesize{\footnotesize}
\centering
\tablewidth{0pt} 
\tablecaption{Abundance Comparision \label{table::averages}} 
\tablehead{ \colhead{Bin} & \colhead{This Work} & \colhead{Caffau+ 2011} & \colhead{Jacobson+ 2014} \\ \colhead{[Fe/H]} & \colhead{$<$[P/Fe]$>$} & \colhead{$<$[P/Fe]$>$} & \colhead{$<$[P/Fe]$>$}  } 
\startdata
0.2 - 0.4 & --0.01 & --0.02 $\pm$ 0.07 & \nodata   \\
0.0 - 0.2 & 0.03 & --0.02 $\pm$ 0.05 & \nodata \\
--0.2 - 0.0 & 0.03 $\pm$ 0.08 & 0.04 $\pm$ 0.09 & --0.14 \\
--0.4 - --0.2 & 0.14 $\pm$ 0.06 & 0.12 $\pm$ 0.10 & --0.17 $\pm$ 0.03 \\
--0.6 - --0.4 & 0.18 $\pm$ 0.13 & 0.32 $\pm$ 0.02 & 0.16 $\pm$ 0.09  \\
\enddata
\tablecomments{Uncertainties are standard deviation from abundances in bin and not due to systematic errors. Numbers without reported standard deviations had a bin size of one star}
\end{deluxetable}

Our measured phosphorus abundances are consistent with the two data sets from \citet{caffau11} and \citet{jacobson14} in the high metallicity range (see Figure \ref{fig::phos_evolution}). Our average [P/Fe] abundance is 0.10 $\pm$ 0.10 dex and the average abundance from \citet{caffau11} is [P/Fe] = 0.08 $\pm$ 0.15 in the same metallicity range. We also found the averages for different metallicity bins are also consistent, as shown in Table \ref{table::averages}.

The abundances of \citet{jacobson14} in the high metallicity range are offset from our abundances near the metallicity [Fe/H] $\sim$--0.2, where their average [P/Fe] measurement is --0.14. Our sample contains six stars between --0.20 $<$ [Fe/H] $<$ 0.25, the average abundance of this sample is 0.04 $\pm$ 0.07 with the lowest abundance for HD 193664 at [P/Fe] = --0.09 $\pm$ 0.08. The UV \ion{P}{1} lines at the high metallicity range saturate and the uncertainty of \citet{roederer14,jacobson14} measurements is $\sim$ 0.2 dex. The offset between the UV and IR sample is within measurement errors in both samples. Our abundance measurements in the higher metallicity bin of --0.60 $<$ [Fe/H] $<$ --0.40 are in close agreement, as shown in Table \ref{table::averages}.
\section{Discussion}
 \label{section::discussion}
   
\subsection{Phosphorus Galactic Chemical Evolution} 
 
\begin{figure*}[t!]
\hspace*{-2cm}
\epsscale{1} 
\centering
\includegraphics[trim=0cm 0cm 0cm 0cm, scale=.33, clip=True]{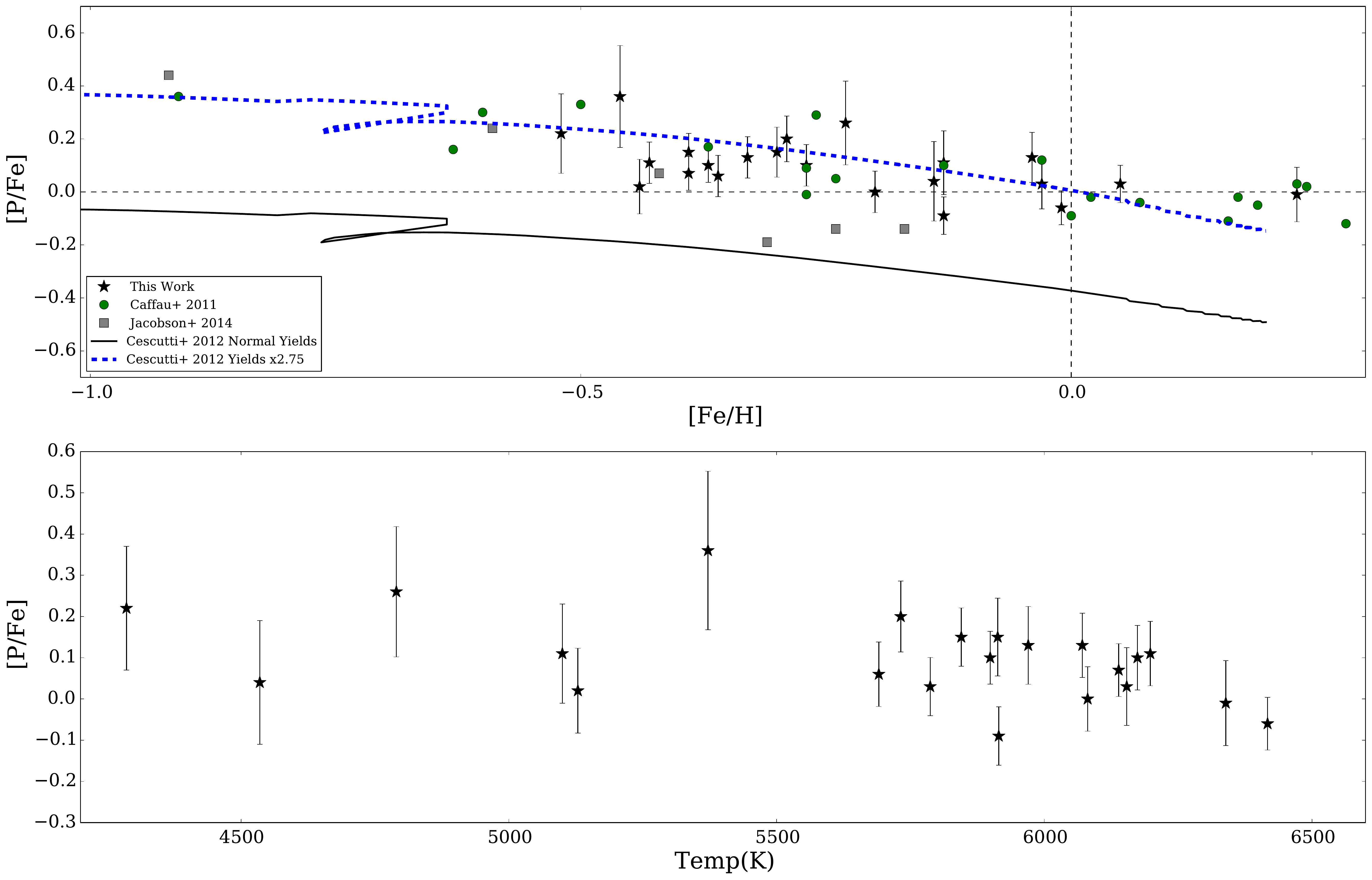} 
\caption{Phosphorus abundances are plotted for our sample (black stars), from \citet{jacobson14}  (grey squares), and from \citet{caffau11} (green circles). The blue dashed line is a chemical evolution model of [P/Fe] in the solar neighborhood with yields increased by 2.75 and the black line is a model with normal yields. These are model 8 and model 6 from \citet{cescutti12} respectively. Dashed lines indicate solar values. \label{fig::phos_evolution}} 
\end{figure*}

We compare our phosphorus abundances, combined with those of \citet{caffau11} and \citet{jacobson14}, to two chemical evolution models from \citet{cescutti12}, shown in Figure \ref{fig::phos_evolution}. The first chemical evolution model uses yields from \citet{kobayashi06} (model 6 from \citealt{cescutti12}) and the second model uses the same yields increased by a factor of 2.75 (model 8 from \citealt{cescutti12}). Additionally, both models include contributions from hypernovae which are necessary to fit measurements at low metallicities but do not significantly contribute to P production at the high metallicity end \citep{jacobson14}. Our results are fit by the enhanced yield model better than the original yield model. To test this, the chemical evolution model's predicted [P/Fe] values were compared to values derived in our sample and abundances from \citet{caffau11}. We found that $\chi ^{2}$ was --17.4 for the model with the original yields and 4.4 for the model with yields increased by 2.75. We therefore find our results are consistent with previous measurements of phosphorus and find chemical evolution models with the predicted yields underproduces P compared to observations. 

\subsection{Phosphorus and the Alpha Elements} 
 \label{subsection::p_alpha}
Additionally, we examined the ratios of [P/Si] and [P/S], the two elements nearest to phosphorus in atomic number.  Available sulfur abundances for our stars have also been taken from \citet{reddy03}. The [S/Fe] ratios from \citet{reddy03} have been adjusted for two effects; first we adjusted their [S/Fe] ratios to account for differences in derived [Fe/H] abundances between our atmospheric parameters from \citet{ramirez13} and those of \citet{reddy03}. Second, the adopted solar abundance from \citet{reddy03} (A(S)=7.34 dex) is different from the adopted solar abundance from \citet{caffau11sulfur} of A(S) = 7.16 dex. We placed the \citet{reddy03} results on the same scale. Finally, we note abundances from the \ion{S}{1} lines measured from multiple 8 at $\sim$ 6750 $\AA$ and at $\sim$ 6050 $\AA$ from multiple 10 used in \citet{reddy03} were calculated using LTE analysis and these lines likely do not suffer from NLTE effects \citep{chen02}.

\begin{figure*}[t!]
\epsscale{1} 
\centering
\includegraphics[trim=4cm 0cm 5cm 0cm, scale=.35, clip=True]{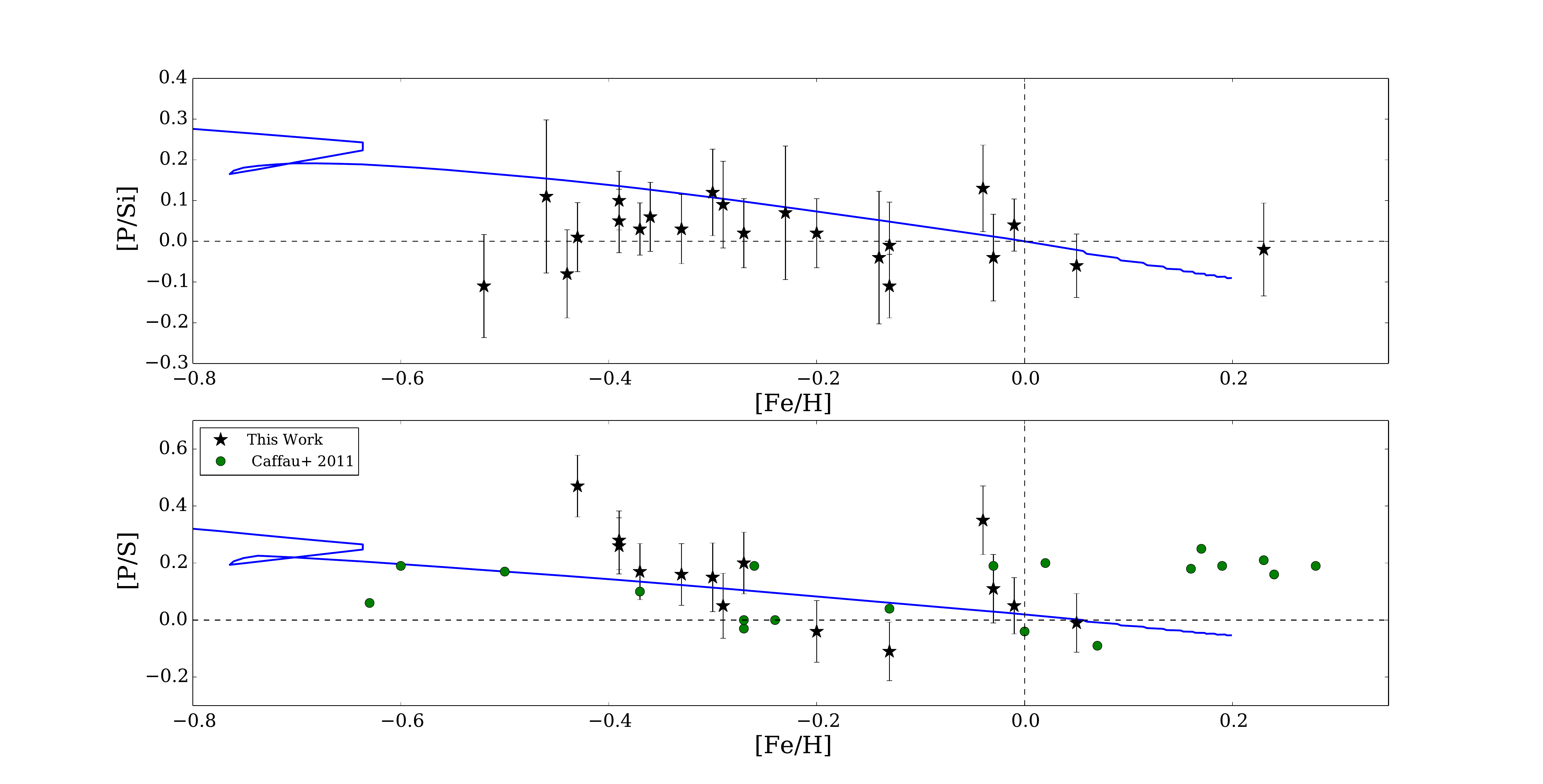} 
\caption{Top Panel: [P/Si] ratios for our sample of stars. Bottom Panel: [P/S] ratios for our stars with sulfur abundances from \citet{reddy03} (black stars), and ratios from \citet{caffau11} (blue circles). Dashed lines indicate solar values. The solid blue line represents a chemical evolution model and is described in section \ref{subsection::p_alpha}. \label{fig::phos_si_s}} 
\end{figure*}

We find that the phosphorus to silicon ratio is consistent with the solar ratio over our metallicity range as shown in Figure \ref{fig::phos_si_s}. The [P/Si] ratio for our abundances is plotted in the upper panel of Figure \ref{fig::phos_si_s} and we find an average ratio of [P/Si]=0.02 $\pm$ 0.07. [P/S] ratios found in \citet{caffau11} are compared to our results plotted in the lower panel of Figure \ref{fig::phos_si_s}. We find an average [P/S] ratio of 0.15 $\pm$ 0.15, in agreement with the result from \citet{caffau11}, who found a ratio of [P/S]=0.10 $\pm$ 0.10 for their sample. Combining our two samples leads to a total ratio of [P/S]=0.13 $\pm$ 0.12. 

Our [P/Si] and [P/S] results are also compared to chemical evolution models in Figure \ref{fig::phos_si_s}. The chemical evolution model for phosphorus was adopted from \citep{cescutti12} and the yields for S was taken from \citet{kobayashi06}. The Si yields were adopted from \citet{francois04} which are the same yields of \citet{woosley95} for solar metallicity. Additionally, the nearly constant [P/Si] and [P/S] ratios over the metallicity range of --0.45 $<$ [Fe/H] $<$ 0.2 may be due insensitivity of the neutron excess that is thought to form the odd elements. The neutron excess increases with increasing metallicity, as more neutron rich material is present in the star. An increase with increasing [Fe/H] is expected because the formation of odd elements like Na and Al is partially due to neutron capture unlike the alpha elements \citep{woosley95}. Our constant [P/S] ratio is in agreement with \citep{caffau11}. However, ratios of [Na,Al/Mg] do not sharply decrease until [Fe/H] $\sim$ --1 \citep{gehren}. The strong decrease with decreasing metallicity in [Na,Al/Fe] ratio is most prominent at [Fe/H] $\sim$ --1 in models as well (e.g. Fig 13 from \citealt{kobayashi11}). Measurements of phosphorus in stars with low metallicities would determine if the neutron excess is important at lower metallicities and if the dependence of [P/Fe] over metallicity is similar to the other odd elements. 

The $^{30}$Si isotope, which is thought to be the seed for P production, has also been measured to be higher than predictions from chemical evolution models. The infrared study of SiO in M giant stars found isotope ratios of $^{28}$Si/$^{30}$Si to be between $\sim$20-30 for their sample of six stars \citep{tsuji94}, while chemical evolution models give the $^{28}$Si/$^{30}$Si $\sim$ 34 at a [Fe/H]= 0 in the solar neighborhood \citep{kobayashi11}. While the predicted ratios are lower than measured by a factor of 1.1 - 1.3, this difference is not extreme. For an A($^{28}$Si)= 7.51, a ratio of 25 would give an A($^{30}$Si)= 6.11 while a ratio of 35 gives an A($^{30}$Si)= 5.97.

\subsection{Possibly Anomalous Stars} 
 
\textit{HD 20794}: This star has a slightly high [P/Fe] ratio of 0.36 $\pm$ 0.18 for a metallicity of [Fe/H] = --0.46 compared to the average [P/Fe] of 0.18 in the metallicity bin --0.6 $<$ [Fe/H] $<$ --0.4 and the average of 0.14 in the metallicity bin --0.4 $<$ [Fe/H] $<$ --0.2 (shown in Table \ref{table::averages}). The high [P/Fe] ratio may be due to uncertain atmospheric parameters; \citet{bensby14} cites the uncertainty in the temperature for this star at $\pm$108K and the uncertainty in the log g of 0.21 \citep{bensby14}. Adding the abundance uncertainties in quadrature for this star gave an uncertainty of 0.17 dex. The lower end of the 1 $\sigma$ uncertainty range would place the phosphorus abundances near other [P/Fe] ratios. The alpha elements for this star are [Mg/Fe] =  0.33 $\pm$ 0.27, [Si/Fe] = 0.22 $\pm$ 0.23, and [Ca/Fe] = 0.13 $\pm$ 0.48 from \citet{bensby14} and this work found a [Si/Fe] = 0.25 $\pm$ 0.12. These high abundances and high velocity for this star make it a probable member of the thick disk \citep{bensby14}. The phosphorus abundance is correlated with the alpha elements in the high metallicity range, as shown in Figure \ref{fig::phos_si_s}, and therefore the [P/Fe] ratio this star may also be high.    
 
\textit{HD 163363}: This star has [P/S] of 0.35 $\pm$ 0.12, which is high when compared to other stars near [Fe/H] $\sim$ 0. The high abundance is due to a low [S/Fe] abundance from \citet{reddy03} rather than to a high phosphorus abundance, as this star has a [P/Fe] = 0.13 $\pm$ 0.09. The low [S/Fe] abundance contrasts with other alpha element abundances in HD 163363: [Mg/Fe] = 0.15 $\pm$ 0.03 and [Si/Fe] = 0.01 $\pm$ 0.05 from \citet{reddy03} and our [Si/Fe] ratio of 0.01 $\pm$ 0.09. The high sulfur abundance may be measurement error, as it is only $\sim$ 3 $\sigma$ from a [P/S]=0, and less than 2 $\sigma$ away from the average value of [P/S] = 0.15 $\pm$ 0.15 found within our sample.  
 
\textit{HD 193664}: The star has a slightly low phosphorus abundance at [P/Fe] = --0.09 $\pm$ 0.07. The other alpha elements are [Mg/Fe] = 0 $\pm$ 0.03, [Si/Fe] = 0.02 $\pm$ 0.05 from \citet{reddy03}, and our [Si/Fe] = 0.02 $\pm$ 0.08. The [P/Fe] abundance is low considering the other alpha elements but it is consistent with the quoted uncertainties; it is 1.7 $\sigma$ away from the average value of [P/Fe] = 0.03 from its metallicity bin (--0.2 $<$ [Fe/H] $<$ 0) in Table \ref{table::averages}. 
 
\textit{HD 194497}: This star has [P/S] = 0.47 $\pm$ 0.11. This is due to a low sulfur abundance, since we find a normal [P/Fe] abundance. The [P/S] is 2.9 $\sigma$ from the average value of [P/S] = 0.15 $\pm$ 0.15. Other alpha abundances are low in this star; [Mg/Fe] = --0.01 $\pm$ 0.03 and [Si/Fe] = 0.03 $\pm$ 0.05 from \citet{reddy03} and our [Si/Fe] = 0.10 $\pm$ 0.08.

\section{Conclusion }
\label{section::conclusion}

\begin{enumerate}

\item{We have derived phosphorus and silicon abundances for 22 stars using infrared spectra observed with Phoenix on the KPNO 4m Mayall telescope, Gemini South Telescope, and for Arcturus using spectra from the infrared Arcturus spectral atlas \citep{hinkle95}.}
\item{We found no systematic difference between [P/Fe] abundances in dwarfs and giants.}
\item{Our phosphorus abundances results are consistent with the other studies, such as \citet{caffau11} and \citet{jacobson14}. We find an average [P/Fe] ratio of 0.10 $\pm$ 0.10 in the metallicity reange --0.55 $<$ [Fe/H] $<$ 0.2. Our results are in agreement with \citet{caffau11}, who found an average abundance of [P/Fe] = 0.08 $\pm$ 0.15 for their sample with  metallicities between --1.0 $<$ [Fe/H] $<$ 0.2. We also compared [P/Fe] averages to \citet{caffau11,jacobson14} in metallicity bins with sizes of 0.2 dex. Our average values in the metallicity bins to \citet{caffau11} agree in all metallicity bins.} 
\item{ Our [P/Fe] measurements do not match the results of chemical evolution models with the most recent nucleosynthesis yields. Instead our results were more consistent with a chemical evolution model of phosphorus with yields increased by a factor of 2.75. }
\item{We measured Si abundances from a nearby \ion{Si}{1} feature. Our silicon abundances matched literature sources; the average difference for 15 stars with corresponding measurements in \citep{reddy03} is --0.03 $\pm$ 0.11 dex. We find an average [P/Si] ratio of 0.02 $\pm$ 0.07 for our sample of stars}
\item{We found a [P/S] ratio of 0.15 $\pm$ 0.15 in our sample of stars over the metallicity range --0.55 $<$ [P/Fe] $<$ 0.2, using S abundances from \citet{reddy03}. This result is consistent with results from \citet{caffau11}, who found [P/S]=0.10 $\pm$ 0.10 over their range from --1 $<$ [P/Fe] $<$ 0.2.}
\item{Other odd light elements, such as Na and Al, are sensitive to the neutron flux and their abundances with respect to the alpha elements decreases. Our constant [P/Si] and [P/S] ratios may imply phosphorus is produced by other processes than neutron capture, however more abundance measurements at lower metallicty ranges, such as [Fe/H] $<$ --1.0 where the decrease in Na and Al with decreasing metallicity are most pronounced, are needed to explore this hypothesis. }

\end{enumerate}

\section{Acknowledgements}
This paper is based on observations obtained at the Kitt Peak National Observatory, a division of the National Optical Astronomy Observatories (NOAO). NOAO is operated by the Association of Universities for Research in Astronomy, Inc. under cooperative agreement with the National Science Foundation. This work is also based on observations obtained at the Gemini Observatory, which is operated by the Association of Universities for Research in Astronomy, Inc., under a cooperative agreement with the NSF on behalf of the Gemini partnership: the National Science Foundation (United States), the National Research Council (Canada), CONICYT (Chile), Ministerio de Ciencia, Tecnolog\'{i}a e Innovaci\'{o}n Productiva (Argentina), and Minist\'{e}rio da Ci\^{e}ncia, Tecnologia e Inova\c{c}\~{a}o (Brazil). The Gemini observations were done under proposal ID GS-2016B-Q-76. We are grateful to the Kitt Peak National Observatory and particularly to Colette Salyk for her assistance at the start of the observing run and to Anthony Paat, Krissy Reetz, and Doug Williams during the run. We thank German Gimeno for his assistance with the Gemini South Telescope observing run. We thank the referee E. Caffau for suggested improvements to the manuscript. This research has made use of the NASA Astrophysics Data System Bibliographic Services, the Kurucz atomic line database operated by the Center for Astrophysics, and the Vienna Atomic Line Database operated at the Institute for Astronomy of the University of Vienna. This research has made use of the SIMBAD database, operated at CDS, Strasbourg, France. This publication makes use of data products from the Two Micron All Sky Survey, which is a joint project of the University of Massachusetts and the Infrared Processing and Analysis Center/California Institute of Technology, funded by the National Aeronautics and Space Administration and the National Science Foundation. We thank Eric Ost for implementing the model atmosphere interpolation code. C. A. P. acknowledges the generosity of the Kirkwood Research Fund at Indiana University. G.C. acknowledges financial support from the European Union Horizon 2020 research and innovation programme under the Marie Sk lodowska-Curie grant agreement No 664931. 

\software{IRAF, MOOG (v2014; \citealt{sneden})}


\end{document}